\documentclass[aps,prl,reprint,superscriptaddress,twocolumn,longbibliography]{revtex4-2}
\usepackage{amsfonts,amssymb,amscd,amsthm}
\usepackage{graphicx}
\usepackage{mathrsfs}
\usepackage{soul,color,xcolor}
\usepackage{amsmath}
\usepackage[colorlinks, citecolor=red]{hyperref}
\usepackage{etoolbox}
\usepackage{upgreek}
\usepackage[T1]{fontenc}

\makeatletter
\let\old@secnumdepth\secnumdepth  
\makeatother
\begin{document}
\title{Long-time storage of entangled logical states in decoherence-free subspaces}

\author{L. Zhang}
\thanks{These authors contribute equally to this work}%
\affiliation{Center for Quantum Information, Institute for Interdisciplinary Information Sciences, Tsinghua University, Beijing 100084, PR China}

\author{Y.-L. Xu}
\thanks{These authors contribute equally to this work}%
\affiliation{Center for Quantum Information, Institute for Interdisciplinary Information Sciences, Tsinghua University, Beijing 100084, PR China}

\author{Y.-K. Wu}

\affiliation{Center for Quantum Information, Institute for Interdisciplinary Information Sciences, Tsinghua University, Beijing 100084, PR China}
\affiliation{Hefei National Laboratory, Hefei 230088, PR China}

\author{C. Zhang}
\affiliation{HYQ Co., Ltd., Beijing 100176, PR China}

\author{Z.-B. Cui}
\affiliation{Center for Quantum Information, Institute for Interdisciplinary Information Sciences, Tsinghua University, Beijing 100084, PR China}

\author{Y.-Y. Chen}
\affiliation{Center for Quantum Information, Institute for Interdisciplinary Information Sciences, Tsinghua University, Beijing 100084, PR China}

\author{W.-Q. Lian}
\affiliation{HYQ Co., Ltd., Beijing 100176, PR China}

\author{J.-Y. Ma}
\affiliation{HYQ Co., Ltd., Beijing 100176, PR China}

\author{B.-X. Qi}
\affiliation{Center for Quantum Information, Institute for Interdisciplinary Information Sciences, Tsinghua University, Beijing 100084, PR China}

\author{Y.-F. Pu}
\affiliation{Center for Quantum Information, Institute for Interdisciplinary Information Sciences, Tsinghua University, Beijing 100084, PR China}
\affiliation{Hefei National Laboratory, Hefei 230088, PR China}

\author{Z.-C. Zhou}
\affiliation{Center for Quantum Information, Institute for Interdisciplinary Information Sciences, Tsinghua University, Beijing 100084, PR China}
\affiliation{Hefei National Laboratory, Hefei 230088, PR China}

\author{L. He}
\affiliation{Center for Quantum Information, Institute for Interdisciplinary Information Sciences, Tsinghua University, Beijing 100084, PR China}
\affiliation{Hefei National Laboratory, Hefei 230088, PR China}

\author{P.-Y. Hou}
\affiliation{Center for Quantum Information, Institute for Interdisciplinary Information Sciences, Tsinghua University, Beijing 100084, PR China}
\affiliation{Hefei National Laboratory, Hefei 230088, PR China}

\author{L.-M. Duan}
\email{lmduan@tsinghua.edu.cn}
\affiliation{Center for Quantum Information, Institute for Interdisciplinary Information Sciences, Tsinghua University, Beijing 100084, PR China}
\affiliation{Hefei National Laboratory, Hefei 230088, PR China}
\affiliation{New Cornerstone Science Laboratory, Institute for Interdisciplinary Information Sciences, Tsinghua University, Beijing 100084, PR China}

\begin{abstract}
The maintenance of quantum entanglement lays the elementary building block of quantum information processing, requiring an integration of long coherence time, sufficient storage capacity, and high-fidelity entangling gates. Here we encode two-qubit entangled states into the decoherence-free subspaces (DFS) of four ions in a cryogenic trap. By crosstalk-free sympathetic cooling under dual-type encoding and multi-state detection which discards the collision-induced leakage error, we achieve a storage lifetime of about one hour for the entangled logical states. We further study the second-order DFS and show its advantage in suppressing the spatially nonuniform noise over the first-order DFS.
Our work paves the way for applications of DFS quantum memories in quantum computing, quantum network and precision measurement.
\end{abstract}

\maketitle

\section{Introduction}
Quantum entanglement marks the fundamental difference between the quantum and the classical worlds. As an essential resource in quantum information science \cite{nielsen2000quantum}, entangled states find wide applications in quantum computing \cite{shor1996fault,gottesman1999demonstrating,oneway}, quantum communication \cite{PhysRevLett.67.661,qss,pra03liar} and quantum metrology \cite{pra96optimal_frequency,sci04enhanced_measure,Toth12pra,PhysRevLett.109.070503}. However, quantum entanglement is fragile and can be easily lost due to the inevitable interaction between the quantum system and the environment. Therefore, the faithful storage of quantum entanglement has become a critical challenge in its applications in various fields. For example, the coherence time of physical qubits sets an intrinsic limit to the fidelity of the elementary entangling gates \cite{DiVincenzo2000,Ladd2010}, and a storage lifetime of entangled states much longer than the communication time is crucial for the efficient scaling of quantum repeaters \cite{PhysRevLett.81.5932,duan2001long,RevModPhys.83.33} and memory-enhanced multipartite entanglement generations \cite{PhysRevLett.97.143601,PhysRevLett.128.080501}.

Ultimately, quantum entanglement can be preserved for arbitrarily long time on a universal quantum computer by encoding into quantum error correction codes and by actively detecting and correcting the errors, as long as the fidelities of the elementary operations are above the fault-tolerant threshold \cite{nielsen2000quantum,Gottesman1998,campbell2017roads}. However, this generally comes with a large overhead in the qubit number and in the gate complexity to entangle the logical qubits \cite{PhysRevA.86.032324,campbell2017roads}. Therefore, for the noisy intermediate-scale quantum (NISQ) devices, hardware-specific techniques have been employed to isolate the system from the environment for various physical platforms like trapped ions \cite{PhysRevLett.95.060502,Haffner2005,PhysRevLett.113.220501,wang2017single,wang2021single,PhysRevA.106.062617,xu2005longtime}, cold atomic ensembles \cite{RevModPhys.75.457,Zhao2009,PhysRevLett.114.050502,yang2016efficient,wang2019efficient,PhysRevX.14.021018}, rare-earth-doped crystals \cite{PhysRevLett.108.190503,PhysRevLett.108.190504,PhysRevLett.108.190505,Lago-Rivera2021,Ortu2022}, and NV centers in diamonds \cite{doi:10.1126/science.1220513,Bar-Gill2013,PhysRevX.9.031045}. At the software level it is also common to use the dynamical decoupling method \cite{PhysRev.80.580,PhysRev.94.630,10.1063/1.1716296,PhysRevLett.98.100504,PhysRevLett.95.180501,PhysRevLett.106.240501} which actively filters out a wide frequency band of noise, and the decoherence-free subspace (DFS) method \cite{PhysRevLett.79.1953,PhysRevLett.81.2594,doi:10.1126/science.1057357} with a passive immunity to global noise.

Among the various physical systems, trapped ions have the advantages of a long coherence time and the high-fidelity and deterministic entangling gates \cite{10.1063/1.5088164,wang2021single,xu2005longtime,hughes2025trappediontwoqubitgates9999}, making it an ideal platform for long-time storage of quantum entanglement. Previously, a coherence time above an hour has been realized for a single ionic qubit in a room-temperature trap \cite{wang2021single}. However, the frequent hopping of ions due to the incessant collisions of background gas molecules prevents the further extension of the storage capacity to accommodate entangled qubits, as the uncontrolled position exchange between ions will destroy the encoded quantum information. A cryogenic trap can solve this problem by largely suppressing the collision-induced ion hopping rate \cite{Pagano_2019}. In this way, a coherence time of hundreds of milliseconds was achieved for above 200 ions \cite{PhysRevA.106.062617}, but there the stored states were written by global microwave pulses and hence encoded no entanglement. Recently Ref.~\cite{xu2005longtime} demonstrated a coherence time above two hours for a logical qubit encoded in the DFS of two physical qubits. In that experiment, two-qubit entangled states $\frac{1}{\sqrt{2}}(|01\rangle \pm |10\rangle)$ were stored, but the memory capacity was still restricted to one single qubit under the protection of the DFS, while other two-qubit states like $\frac{1}{\sqrt{2}}(|00\rangle \pm |11\rangle)$ got larger sensitivity to noise than a physical qubit \cite{xu2005longtime}. Therefore, a combination of a long storage lifetime, a capacity above one logical qubit, and a universal gate set to encode multi-qubit quantum information, remains an experimental challenge.

In this work, we encode two logical qubits into the DFS of four memory ions in a cryogenic trap, and report the storage of logical entangled states with a lifetime of about one hour. We demonstrate the combination of dual-type qubit encoding which allows us to perform crosstalk-free sympathetic laser cooling \cite{yang2022realizing,10.1063/5.0069544}, individually addressed high-fidelity entangling gates which enables the preparation of arbitrary logical states, and the multi-state detection technique which separates the leakage error from other storage infidelities \cite{xu2005longtime}. We also note that a two-qubit entangled DFS state $\frac{1}{\sqrt{2}}(|1001\rangle\pm|0110\rangle)$ corresponds to a second-order DFS state against spatially inhomogeneous noise. We hence compare its storage performance with a first-order DFS state, and show that the higher-order DFS can better suppress the nonuniform magnetic field noise, thus further improving the storage lifetime.

\begin{figure*}[!tp]
   \includegraphics[width=\linewidth]{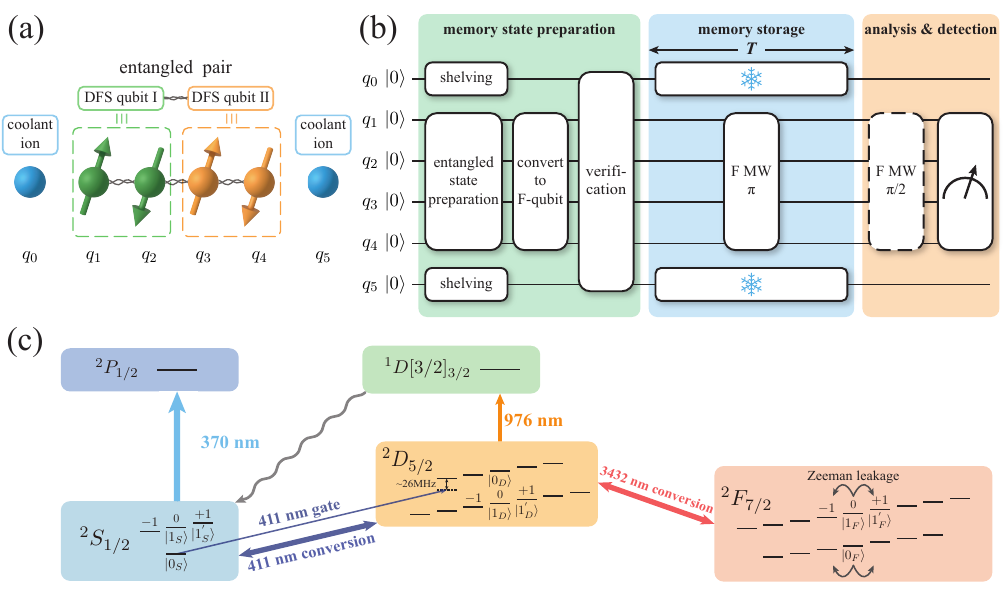}
   \caption{Experimental Scheme. (a) Six ${}^{171}\mathrm{Yb}^+$ ions form a one-dimensional chain, with two edge ions (blue) for sympathetic cooling and four central ions encoding two logical qubits (green and orange) in decoherence-free subspaces (DFS).
   (b) Experimental sequence. We initialize the ions in $|0_S\rangle$, and use focused $411\,$nm laser beams and a global microwave to prepare the entangled states between the four memory ions in the DFS. The memory ions are then transferred to the $F$-type for long-time storage, while the coolant ions are not affected by shelving into an ancilla level. After a verification step which checks the correct qubit types, the $F$-type memory qubits are stored for time $T$ with a microwave spin echo inserted in the middle, while the $S$-type coolant ions continuously provide sympathetic laser cooling. Finally, the memory ions undergo analysis pulses and multi-state detection to measure the storage fidelity and the leakage error.
   (c) Relevant level structure of the ${}^{171}\mathrm{Yb}^+$ ion. Dual-type qubits are defined by the clock states of $S_{1/2}$ ($|0_S\rangle$ and $|1_S\rangle$) and $F_{7/2}$ ($|0_F\rangle$ and $|1_F\rangle$) levels, with their coherent conversion achieved by global bichromatic $411\,$nm and $3432\,$nm laser beams. The population on the $F$-type qubits may slowly leak to the nearby Zeeman levels due to the collision of the ions with the background $H_2$ gas molecules. We employ global $370\,$nm laser for laser cooling, optical pumping and fluorescence detection, $976\,$nm repump laser to clear the $D_{5/2}$ states, and focused $411\,$nm laser beams for single-qubit and two-qubit gates.
   \label{fig1}}
\end{figure*}

\section{Experimental scheme}
As illustrated in Fig.~\ref{fig1}(a), we trap a linear chain of six ${}^{171}\mathrm{Yb}^+$ ions in a
cryogenic trap at a temperature of $6\,$K. The four central ions form two neighboring pairs labeled by green and orange colors to encode two logical qubits in their decoherence-free subspaces (DFS), while the two edge ions provide sympathetic cooling to maintain the stability of the whole chain during the storage.
The experimental sequence is shown in Fig.~\ref{fig1}(b) with the relevant energy levels of the ions sketched in Fig.~\ref{fig1}(c). We use the dual-type qubit scheme to encode the memory and the coolant ions \cite{yang2022realizing}. The memory ions are initialized in the $S$-type states $|0_S\rangle \equiv |S_{1/2},F=0,m_F=0\rangle$ and $|1_S\rangle \equiv |S_{1/2},F=1,m_F=0\rangle$ and are prepared into the two-qubit DFS logical states spanned by $|0_{LS}\rangle\equiv|1_S 0_S\rangle$ and $|1_{LS}\rangle\equiv|0_S 1_S\rangle$ through a sequence of single-qubit and two-qubit gates using focused $411\,$nm laser beams and $12.6\,$GHz global microwave pulses. Then we convert the memory ions to the $F$-type states spanned by $|0_F\rangle \equiv |F_{7/2},F=3,m_F=0\rangle$ and $|1_F\rangle \equiv |F_{7/2},F=4,m_F=0\rangle$ for long-time storage using global bichromatic $411\,$nm and $3432\,$nm laser beams, while the coolant ions are not affected as we temporarily shelve them to the ancilla level $|D_{5/2},F=2,m_F=-1\rangle$ using focused $411\,$nm laser. The successful qubit-type conversion can be verified by all the memory ions being dark while both coolant ions being bright under the global $370\,$nm cooling laser and the $976\,$nm and the $935\,$nm (not shown in Fig.~\ref{fig1}) repump lasers. Then after a storage time $T$ with a $3.6\,$GHz microwave spin echo in the middle on the $F$-type qubits, we determine the storage fidelity by measuring in different bases. During the measurement, apart from the desired qubit states $|0_F\rangle$ and $|1_F\rangle$, we also distinguish the leakage error to the $m_F\ne 0$ Zeeman levels of the $F_{7/2}$ manifolds which likely occurs due to the collision of the ions with the background $H_2$ gas molecules \cite{xu2005longtime}. More details can be found in Appendix A and in Ref.~\cite{xu2005longtime}.

\section{Encoding two-qubit entangled states in DFS}
\begin{figure}[!tbp]
   \includegraphics[width=\linewidth]{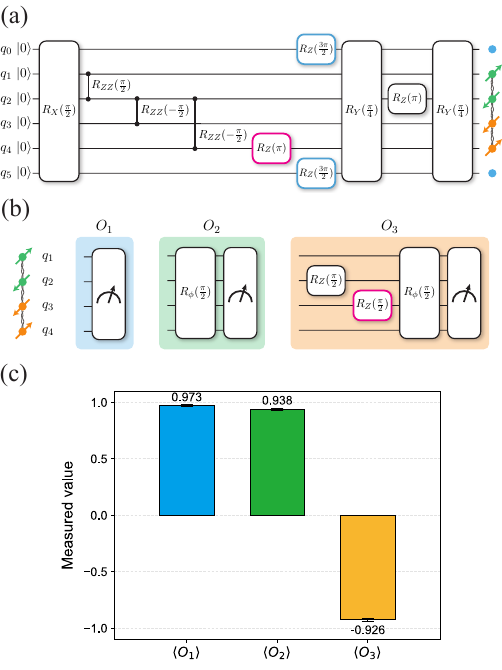}
   \caption{Entanglement of DFS-encoded qubits.
   (a) Quantum circuit to prepare the entangled state $|\psi^{+}_L\rangle\equiv\frac{1}{\sqrt{2}}(|0_L 1_L\rangle + |1_L 0_L\rangle)=\frac{1}{\sqrt{2}}(|1001\rangle+|0110\rangle)$. $R_X$ and $R_Y$ rotations are achieved by global microwave pulses and hence also act undesirably on the coolant ions, which will be compensated by the single-qubit $R_Z(3\pi/2)$ rotations indicated by the blue boxes. The $R_{ZZ}(\pm\pi/2)\equiv e^{\mp i (\pi/4)Z\otimes Z}$ and $R_Z$ gates are realized by focused $411\,$nm laser beams. (b) The fidelity of the GHZ-like entangled state is decomposed into three terms $F=\frac{1}{4}(2\langle O_{1}\rangle + \langle O_{2}\rangle - \langle O_{3}\rangle)$ which are measured separately by the three circuits.
   (c) Fidelity of the prepared $|\psi^{+}_L\rangle$ state without conversion to $F$-type or storage. A fidelity $F=95.3(5)\%$ is obtained.
   \label{fig2}}
\end{figure}

The two-qubit entangled logical states in the DFS $|\phi^{\pm}_L\rangle\equiv\frac{1}{\sqrt{2}}(|0_L 0_L\rangle \pm |1_L 1_L\rangle)=\frac{1}{\sqrt{2}}(|1010\rangle \pm |0101\rangle)$ and $|\psi^{\pm}_L\rangle \equiv \frac{1}{\sqrt{2}}(|0_L1_L\rangle \pm|1_L0_L\rangle)=\frac{1}{\sqrt{2}}(|1001\rangle \pm |0110\rangle)$ are GHZ-like states and can be prepared by the quantum circuits in Fig.~\ref{fig2}(a). Here we take the state $|\psi^{+}_L\rangle$ as an example, which can be decomposed into three maximally entangled two-qubit gates together with a few single-qubit rotations. Specific to our system, we have individually addressed two-qubit light shift gates $R_{ZZ}(\pm\pi/2)= e^{\mp i(\pi/4)Z\otimes Z}$ \cite{xu2005longtime,PhysRevA.103.012603} with an average Bell state fidelity of $99.1\%$ between arbitrary memory ion pairs (see Appendix B) and single-qubit $Z$ rotations $R_{Z}(\theta)= e^{-i(\theta/2)Z}$ through focused $411\,$nm laser beams whose orientations are controlled by acousto-optic deflectors. As for the single-qubit rotations around the $X$ or $Y$ directions, they are achieved by global microwave pulses and hence act collectively on all the qubits, including the four central memory ions ($q_1$-$q_4$) and the two edge coolant ions ($q_0$ and $q_5$). Therefore, we perform two additional single-qubit $R_Z(3\pi/2)$ rotations, indicated by the blue boxes in Fig.~\ref{fig2}(a), to compensate the effect of the global microwave pulses and to ensure that the coolant ions remain in $|0_S\rangle$ after the whole circuit. Then we can further shelve the coolant ions to $|D_{5/2},F=2,m_F=-1\rangle$ and convert the memory ions to the $F$-type as described in Fig.~\ref{fig1}(b).
Also note that, in the implementation of the light shift gates $R_{ZZ}(\pm \pi/2)$, we insert a microwave spin echo in the middle to compensate the single-qubit AC Stark shifts, therefore the actually achieved gates should be multiplied by $\otimes_{i=0}^5 Y_i$ (see Appendix B and Ref.~\cite{xu2005longtime} for more details). Fortunately, the circuit is designed in such a way that these additional $Y$ gates do not change the output state $|\psi^{+}_L\rangle$.

The other logical Bell states can be prepared in similarly ways. Specifically, to get $|\phi^{+}_L\rangle$  we simply move the $R_Z(\pi)$ gate on $q_4$ in Fig.~\ref{fig2}(a) (the red box) to $q_3$, while $|\psi^{-}_L\rangle$ and $|\phi^{-}_L\rangle$ can be obtained by adding an $R_Z(\pi)$ gate to $|\psi^{+}_L\rangle$ and $|\phi^{+}_L\rangle$, respectively. Since the single-qubit $R_Z$ gates have much lower infidelities than the two-qubit entangling gates, in the following we only calibrate the preparation fidelity of the $|\psi^{+}_L\rangle$ state, and expect similar performance for the other logical Bell states.

To determine the fidelities of the GHZ-like states, we decompose them into three observables \cite{PhysRevA.76.030305,PhysRevLett.94.060501} which can be measured by the three circuits in Fig.~\ref{fig2}(b). We have $|\psi_L^+\rangle\langle\psi_L^+|=\frac{1}{4}(2O_1+O_2-O_3)$ with
\begin{equation}
\begin{aligned}
O_1 \equiv \frac{1}{8}\big( & IIII+ZZZZ+IZZI+ZIIZ \\
                       & - ZIZI -IZIZ -ZZII -IIZZ \big),
\label{eq:1}
\end{aligned}
\end{equation}
\begin{equation}
O_2 \equiv \frac{1}{2\pi} \int_{0}^{2\pi} d\phi \left(X\cos \phi + Y \sin \phi \right)^{\otimes4},
\label{eq:2}
\end{equation}
and
\begin{equation}
\begin{aligned}
O_3 \equiv \frac{1}{2\pi} \int_{0}^{2\pi}d\phi \big[& \left(X\cos \phi + Y \sin \phi \right) \\
& \otimes\left(-X\sin \phi + Y \cos \phi\right) \\
& \otimes\left(-X\sin \phi + Y \cos \phi\right) \\
& \otimes\left(X\cos \phi + Y \sin \phi\right)\big].
    \label{eq:3}
\end{aligned}
\end{equation}
Indeed, as we show in more details in Appendix C, $O_1$ corresponds to the population while $(O_2-O_3)/2$ gives the parity of the targeted GHZ-like state.

The expectation value of $O_1$ can easily be obtained by measuring all the qubits in the computational basis as in the left panel of Fig.~\ref{fig2}(b). To evaluate $O_2$, we apply a global microwave $\pi/2$ pulse with its phase $\phi$ uniformly sampled between $0$ and $2\pi$, followed by $Z$ basis measurements as shown in the middle panel. As for $O_3$, we perform similar $\pi/2$ rotations with random phases, but we need an additional $\pi/2$ relative phase for the rotations on $q_1$, $q_4$ and $q_2$, $q_3$, which can be achieved by the preceding $R_Z(\pi/2)$ gates on $q_2$, $q_3$ as shown in the right panel. In this way, we obtain the measurement outcomes in Fig.~\ref{fig2}(c) which give a preparation fidelity $F=95.3(5)\%$ of the logical Bell state $|\psi^{+}_L\rangle$.
Similarly, to measure the fidelity of $|\phi^{+}_L\rangle$, we can employ almost the same circuits with the only difference being the $R_Z(\pi/2)$ gate acting on $q_3$ (colored in red in the right panel) moved to $q_4$. Note that here to measure the state preparation fidelity, we do not convert the memory ions to the $F$-type for storage, and directly perform rotations on the $S$-type qubits.

\section{Long-time storage of entangled DFS states}
Next, we measure the storage lifetime of the two-qubit entangled DFS states. Observe that the circuits to prepare the states $|\psi^+_L\rangle$ and $|\psi^-_L\rangle$ ($|\phi^+_L\rangle$ and $|\phi^-_L\rangle$) only differ by a single-qubit $Z$ gate which commutes with the dominant storage errors such as the leakage to other Zeeman levels or the dephasing. Therefore, here to save the experimental time cost, we focus on the storage performance of the $|\psi^+_L\rangle$ and the $|\phi^+_L\rangle$ states, and expect similar lifetimes of the other two logical Bell states.

As described in Fig.~\ref{fig1}(b), after the state preparation and the successful verification, we convert the memory ions to the $F$-type for a tunable storage time $T$. Then we measure the storage fidelity directly for the $F$-type qubits rather than first mapping them back to the $S$-type, so that we can use our previously developed multi-state detection technique to identify the possible leakage errors at the same time (see Appendix A and Ref.~\cite{xu2005longtime} for more details). The measurement circuits are similar to those for the $S$-type qubits in Fig.~\ref{fig2}(b), but one difference is that for the $F$-type qubits we do not have direct single-qubit $R_Z$ gates. Therefore, to evaluate the expectation value of $O_3$, we choose to move the required $R_Z(\pi/2)$ gates all the way back to the state preparation stage. This is valid because as we mention above, these $R_Z$ gates commute with the dominant memory errors.

\begin{figure}[!tbp]
   \includegraphics[width=\linewidth]{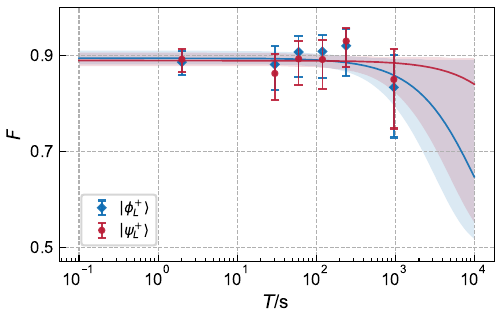}
   \caption{Storage fidelity $F$ versus storage time $T$ for two DFS-encoded Bell states $|\psi^{+}_L\rangle=\frac{1}{\sqrt{2}}(|1001\rangle+|0110\rangle)$ (red circles) and $|\phi^{+}_L\rangle=\frac{1}{\sqrt{2}}(|1010\rangle+|0101\rangle)$ (blue diamonds) after discarding the leakage events. The solid lines represent exponential fitting results, with the shaded regions showing $68\%$ confidence intervals.
   \label{fig3}}
\end{figure}

The experimental results are presented in Fig.~\ref{fig3}. After discarding the leakage events, we obtain $M=250$, $80$, $70$, $60$, $50$, $30$ samples for each stored state and each observable ($O_1$, $O_2$ or $O_3$) at the storage times $T=2\,$s, $30\,$s, $60\,$s, $120\,$s, $240\,$s and $960\,$s, respectively. On the one hand, due to the long storage lifetime of our quantum memory, only a small decrease in the storage fidelity is observed even after $T=960\,$s. On the other hand, for such a long storage time we are only able to collect tens of samples for experimental efficiency, resulting in relatively large error bars. Therefore, when fitting the experimental data with an exponential decay $F=(1+Ae^{-T/\tau})/2$ by the maximum likelihood method, we choose to report the lower bounds of the $68\%$ confidence interval to estimate the storage lifetimes $\tau_{\mathrm{lower}}(|\psi^{+}_L\rangle)=4.9\times10^{3}\,$s and $\tau_{\mathrm{lower}}(|\phi^{+}_L\rangle)=3.2\times10^{3}\,$s, as the upper bounds are almost infinity.

\section{Storage performance of higher-order DFS}

\begin{figure}[!tbp]
   \includegraphics[width=\linewidth]{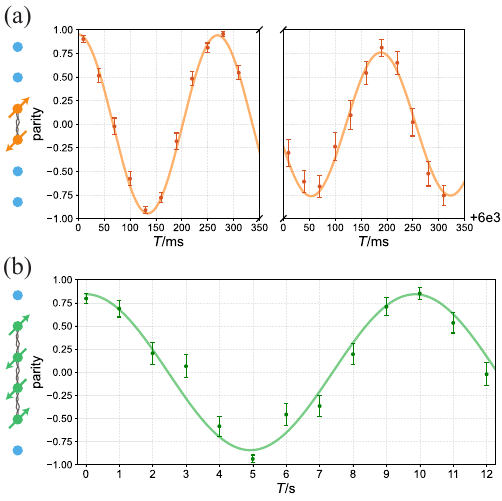}
   \caption{Parity oscillation for the first- and the second-order DFS-encoded states on magnetic-field-sensitive levels. (a) A first-order DFS state $|+_{L1}\rangle=\frac{1}{\sqrt{2}}(|1^\prime0\rangle+|01^\prime\rangle)$ encoded on two central ions. The parity dynamics are measured for two $350$-ms time windows starting from $T=0$ (left panel) and $T=6\,$s (right panel), respectively. (b) A second-order DFS state $|+_{L2}\rangle\equiv|\psi^{+}_L\rangle=\frac{1}{\sqrt{2}}(|1^\prime001^\prime\rangle+|01^\prime1^\prime0\rangle)$ encoded on four central ions.
   \label{fig4}}
\end{figure}

Although the longer storage lifetime of $|\psi^{+}_L\rangle$ than $|\phi^{+}_L\rangle$ may be explained by the statistical fluctuation, we note that $|\psi^{+}_L\rangle=\frac{1}{\sqrt{2}}(|1001\rangle+|0110\rangle)$ locates in the second-order DFS of four ions, while $|\phi^{+}_L\rangle=\frac{1}{\sqrt{2}}(|1010\rangle+|0101\rangle)$ belongs to the first-order DFS. Specifically, while both $|\psi^{+}_L\rangle$ and $|\phi^{+}_L\rangle$ are insensitive to global noise, the $|\psi^{+}_L\rangle$ state possesses an additional robustness to the first-order spatial nonuniformity, if we assume a symmetric configuration of the ion chain.
Here we illustrate the difference between the first-order and the second-order DFS states by purposely enlarging the sensitivity to the magnetic field noise and by removing the spin echo in the storage sequence. We encode the qubits in the magnetic-field-sensitive levels between $|0_S\rangle$ and $|1^\prime_S\rangle \equiv |S_{1/2},F=1,m_F=+1\rangle$, and between $|0_F\rangle$ and $|1^\prime_F\rangle \equiv |F_{7/2},F=4,m_F=+1\rangle$, with their coherent conversion intermediated by $|0_D\rangle$ and $|1_D^\prime\rangle\equiv |D_{5/2},F=3,m_F=+1\rangle$ as shown in Fig.~\ref{fig1}(c). We plot the parity for the first-order and the second-order DFS states in Fig.~\ref{fig4}, encoded in two and four central ions, respectively. For the first-order DFS state $|+_{L1}\rangle=\frac{1}{\sqrt{2}}(|1^\prime0\rangle+|01^\prime\rangle)$, we observe an oscillation period of $269.1(1)\,$ms due to a constant magnetic field difference between the two ions, and a decay of the oscillation amplitude at the timescale of $27.6(81)\,$s due to the time-dependent magnetic field noise. On the other hand, for the second-order DFS state $|+_{L2}\rangle\equiv|\psi^{+}_L\rangle=\frac{1}{\sqrt{2}}(|1^\prime001^\prime\rangle+|01^\prime1^\prime0\rangle)$, we get a much longer oscillation period of $9.9(1)\,$s, and observe no significant decay in the oscillation amplitude on the measured timescale of $12\,$s. Both these results showcase the superior storage performance of the higher-order DFS against spatially inhomogeneous noise.

\section{Discussion}
In this work, we demonstrate the long-time storage of entangled logical states in the four-dimensional DFS of four physical qubits. Instead of first preparing a logical product state like $|0_L 0_L\rangle=|1010\rangle$ and then performing a logical entangling gate, we directly compile the entanglement preparation circuit into the native gate set of global $R_X$ and $R_Y$ gates, and individually addressed $R_Z$ and $R_{ZZ}$ gates. However, note that this native gate set is already universal, as the individual rotation $R_X(\theta)$ can be implemented by $R_Z(\theta)$ sandwiched between global $R_Y(\pi/2)$ and $R_Y(-\pi/2)$ rotations. Therefore, our work can be directly generalized to longer ion chains encoding more DFS logical qubits, thus supporting the storage of larger entangled states. Also note that a trapped ion chain supports distant entangling gates such as the $R_{ZZ}(-\pi/2)$ between $q_2$ and $q_4$ which we demonstrate in Fig.~\ref{fig2}(a). Hence as the number of memory ions increase, we can place more coolant ions in the middle to maintain the efficiency of sympathetic cooling \cite{Lin2016,PhysRevLett.127.143201}, without hampering the capability of entangling different memory ions. Besides, the use of the dual-type qubit scheme ensures the same mass of the memory and the coolant ions, which is also beneficial for sympathetic cooling \cite{PhysRevA.103.012610}. Finally, by adding more memory ions, it is also possible to enhance the storage lifetime by encoding into higher-order DFS which has stronger robustness against spatial inhomogeneity of the noise. However, this needs to be balanced with the increased state preparation error for encoding. Besides, encoding into higher-order DFS also means an increase in the spatial extension of the logical qubits, which will enlarge the spatial fluctuation they may experience and will compete with the lowered sensitivity to noise.

\bigskip
\textbf{Data Availability:} The data that support the findings of this work are available at \cite{data}.

\textbf{Acknowledgements:} This work was supported by Quantum Science and Technology-National Science and Technology Major Project (Grant No. 2021ZD0301601), Beijing Science and Technology Planning Project (Grant No. Z25110100040000), the National Natural Science Foundation of China (Grant No. 12575021 and 12574541), Tsinghua University Initiative Scientific Research Program, and the Ministry of Education of China. L.M.D. acknowledges in addition support from the New Cornerstone Science Foundation through the New Cornerstone Investigator Program. Y.K.W., Y.F.P. and P.Y.H. acknowledge in addition support from the Dushi program from Tsinghua University.

\setcounter{secnumdepth}{0}

\section{Appendix A: Multi-state detection scheme}
\begin{figure*}[!tp]
    \includegraphics[width=\linewidth]{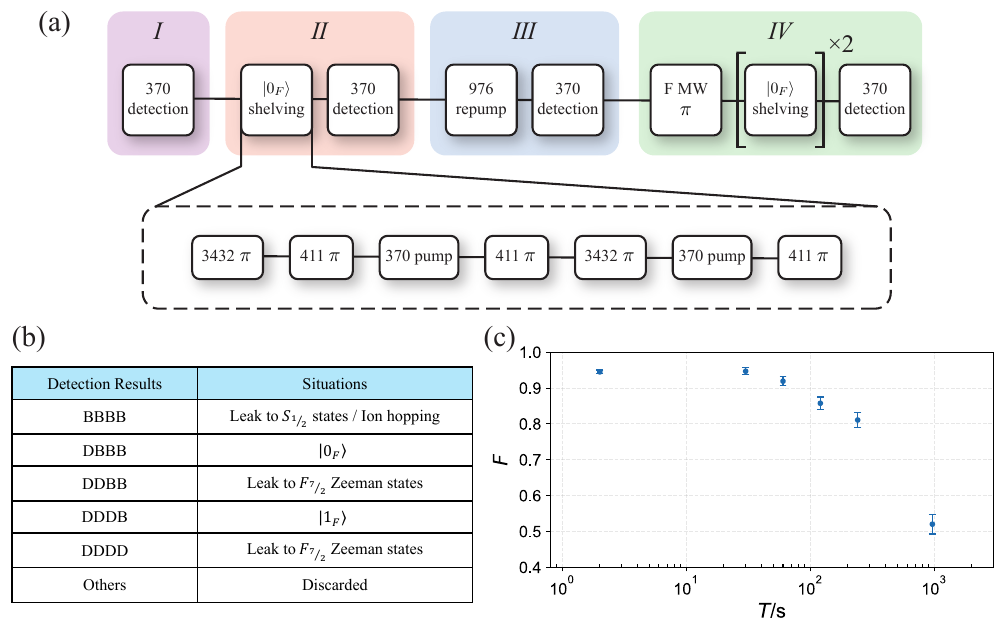}
    \caption{(a) Experimental sequence for multi-state detection of $F_{7/2}$ levels. (b) Lookup table for the measurement outcome (``D'' for dark and ``B'' for bright). (c) Measured survival probability for the four-ion state versus storage time.}
    \label{fig:S2}
\end{figure*}

We follow the multi-state detection scheme in our previous work \cite{xu2005longtime} to distinguish the $|0_F\rangle$, $|1_F\rangle$ and other leakage states. The experimental sequence is illustrated in Fig.~\ref{fig:S2}(a) and is divided into four parts, each involving a fluorescence detection of $S_{1/2}$ under a global $370\,$nm laser beam.
The first stage performs a direct fluorescence detection to identify any leakage events back to $S_{1/2}$ or the possible position exchange between the memory ions and the coolant ions due to the collision with the background gas molecules.
The stage II involves a global bichromatic $3432\,$nm laser (converting $|0_F\rangle$ and $|1_F\rangle$ to $|0_D\rangle$ and $|1_D\rangle$), a monochromatic $411\,$nm laser (converting $|0_D\rangle$ to $|0_S\rangle$) and a $370\,$nm pumping laser to selectively shelve $|0_F\rangle$ to the
$S_{1/2},F=1$ levels. As shown in the dashed box, we further perform a $411\,$nm $\pi$ pulse to clear the residual population on $|0_D\rangle$ due to the pulse imperfection, and a $3432\,$nm $\pi$ pulse to suppress the population in $|0_F\rangle$ while recovering the population in $|1_F\rangle$ from $|1_D\rangle$. Then we apply the final $370\,$nm pumping laser and the $411\,$nm $\pi$ pulse to shelve the newly generated population in $|0_D\rangle$ from $|0_F\rangle$ into $S_{1/2}$ levels. The fluorescence under the followed $370\,$nm detection laser thus indicates the original state $|0_F\rangle$.
In this way, we achieve a shelving detection fidelity $F_{0}=99.6(1)\%$ for the $|0_F\rangle$ state.

If originally the ion has leaked to $m_F\ne 0$ Zeeman levels of $F_{7/2}$, it may be excited off-resonantly to the corresponding $D_{5/2}$ Zeeman levels by the $3432\,$nm $\pi$ pulses in the stage II. Therefore in the third stage we clear the population in $D_{5/2}$ by a $976\,$nm repump laser and then apply a $370\,$nm laser to partially detect the leakage error.
Finally we convert the other qubit state $|1_F\rangle$ into $|0_F\rangle$ using a microwave $\pi$ pulse, and further to the $S_{1/2}$ levels by two rounds of $|0_F\rangle$ shelving sequence in the stage II. The resulting bright state under $370\,$nm detection laser will indicate an original state $|1_F\rangle$, while a final dark state again represents the leakage error to the other $F_{7/2}$ Zeeman levels. In this way, we achieve a detection fidelity $F_1=98.1(2)\%$ for the $|1_F\rangle$ state. Note that both detection fidelities are improved compared with our previous work \cite{xu2005longtime} because of the upgraded frequency locking system for the $3432\,$nm laser.

In Fig.~\ref{fig:S2}(c) we plot the survival probability of the four-ion states versus the storage time from the experimental data in Fig.~\ref{fig3}. In particular, we note that at the storage time $T=960\,$s, about half of the experimental data are discarded due to the leakage of at least one out of four memory ions. This is consistent with our previous results of about $12\%$ leakage error per ion under $800\,$s storage.

\section{Appendix B: Realization of two-qubit entangling gate}
\begin{figure*}[!tp]
    \includegraphics[width=\linewidth]{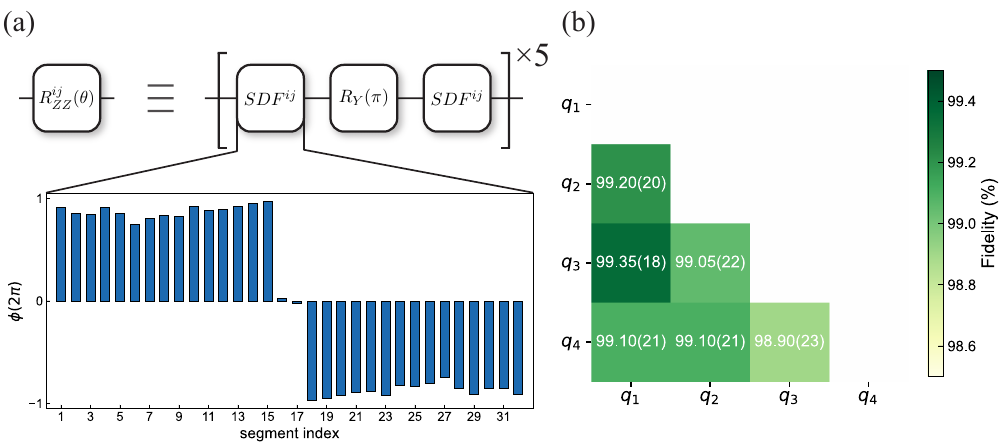}
    \caption{(a) Experimental sequence to realize two-qubit entangling gates between arbitrary memory ion pairs. (b) Bell state fidelity between all the memory ion pairs.}
    \label{fig:S1}
\end{figure*}

We use the experimental sequence in Fig.~\ref{fig:S1}(a) to realize the two-qubit entangling gate $R^{ij}_{ZZ}(\theta)=e^{-i(\theta/2)Z_i Z_j}$ between arbitrary memory ion pairs \cite{xu2005longtime}. We use focused counter-propagating $411\,$nm laser beams, red detuned by $\Delta\pm\mu/2$ ($\Delta=2\pi\times 26\,$MHz) from the $|0_S\rangle\leftrightarrow|D_{5/2},F=2,mF=-2\rangle$
transition, to realize a spatial-varying AC Stark shift on the $|0_S\rangle$ state, namely a spin-dependent force (SDF). Intermediated by the collective spatial oscillation of the ions, this leads to a $|0_S\rangle_i\langle0_S|\otimes |0_S\rangle_j\langle0_S|$ interaction. By further performing a microwave $\pi$ pulse between two SDF sequences, we thus obtain the $Z_i Z_j$ entanglement while cancelling the net AC Stark shift on individual ions between the two pulses.
Specifically, for our four memory ions in the six-ion chain with transverse phonon mode frequencies of $2\pi\times(1.303,1.347,1.385,1.416,1.441,1.458)\,$MHz, we set the beat note frequency of the $411\,$nm laser to be $\mu=2\pi\times1.337\,$MHz. For different ion pairs, we choose a SDF sequence length of $140$-$160\,\upmu$s and divide it into $22$-$36$ equal segments to decouple the spin and the phonon states by phase modulation \cite{PhysRevLett.114.120502}. (An example is shown in Fig.~\ref{fig:S1} for the two central memory ions.) We use an antisymmetric phase sequence to enhance the robustness against the slow drift of the trap frequency \cite{PhysRevLett.120.020501}, and we divide the whole $R_{ZZ}(\pm \pi/2)$ gate into five $R_{ZZ}(\pm \pi/10)$ gates to lower the required laser intensity and to suppress the uncompensated AC Stark shift. In this way, we achieve the Bell state fidelity between arbitrary memory ion pairs in Fig.~\ref{fig:S1}(b), with an average value of $F=99.1\%$.

\makeatletter
\renewcommand{\theequation}{C\arabic{equation}}
\makeatother
\setcounter{equation}{0}

\section{Appendix C: Fidelity of four-ion GHZ-like states}
Here we describe the decomposition of the fidelity of the GHZ-like states, and the physical meanings of individual terms.

We take $|\psi^{+}_L\rangle=\frac{1}{\sqrt{2}}(|1001\rangle+|0110\rangle)$ as an example, whose density matrix can be expanded into
\begin{equation}
\begin{aligned}
    |\psi^{+}_L\rangle\langle\psi^{+}_L| =& \frac{1}{2}\left(|1001\rangle\langle1001|+|0110\rangle\langle0110|\right) \\
    &+\frac{1}{2}\left(|1001\rangle\langle0110|+|0110\rangle\langle1001|\right) \\
    =&\frac{IIII+ZZZZ+IZZI+ZIIZ}{16} \\
    &-\frac{ZIZI+IZIZ+ZZII+IIZZ}{16} \\
    &+\frac{XXXX+YYYY-XYYX-YXXY}{16} \\
    &+\frac{XYXY+YXYX+XXYY+YYXX}{16}.
    \label{eq:4}
\end{aligned}
\end{equation}
The first term is just $O_1/2$ defined in the main text and describes the population in $|1001\rangle\langle1001|+|0110\rangle\langle0110|$. It can be obtained directly by measuring all the memory ions in the $Z$ basis. On the other hand, the second term involves measurements in the $X$ and the $Y$ bases, which can be deduced from a global analysis $\pi/2$ pulse with a phase of $\phi$. For such a $\pi/2$ pulse along the direction of $-X\sin\phi+Y\cos\phi$, we can get $(X\cos\phi+Y\sin\phi)^{\otimes4}$. By further averaging over random $\phi$ distributed uniformly on $[0,2\pi]$ and taking the expectation value, we get
\begin{equation}
\begin{aligned}
\langle O_2\rangle =& \frac{1}{2\pi}\int_{0}^{2\pi}d\phi\bigl\langle(X\cos\phi+Y\sin\phi)^{\otimes4}\bigr\rangle \\
    =& \frac{3}{8}\langle XXXX+YYYY\rangle+\frac{1}{8}\langle XYXY+YXYX \\
    &+XXYY+YYXX+XYYX+YXXY\rangle.
    \label{eq:5}
\end{aligned}
\end{equation}
It covers the same Pauli strings as the second term in Eq.~(\ref{eq:4}), but with different coefficients. To recover all the desired terms, we further introduce a third observable with a phase shift of $\pi/2$ for the analysis pulses on the two central ions. This gives
\begin{equation}
\begin{aligned}
\langle O_3\rangle =& \frac{1}{2\pi}\int_{0}^{2\pi}d\phi\bigl\langle(X\cos\phi+Y\sin\phi) \\
      & \otimes(-X\sin\phi+Y\cos\phi) \\
      & \otimes(-X\sin\phi+Y\cos\phi) \\
      & \otimes(X\cos\phi+Y\sin\phi)\bigr\rangle \\
    =& \frac{1}{8}\langle XXXX+YYYY\rangle-\frac{1}{8}\langle XYXY+YXYX \\
    &+XXYY+YYXX\rangle+\frac{3}{8}\langle XYYX+YXXY\rangle.
    \label{eq:6}
\end{aligned}
\end{equation}
Combining them together, we get the parity term $|1001\rangle\langle0110|+|0110\rangle\langle1001|=(O_2-O_3)/2$.

%

\end{document}